\documentclass[lettersize,journal]{IEEEtran}
\usepackage{amsmath,amsfonts}
\usepackage[noend]{algorithmic}
\usepackage{algorithm}
\usepackage{array}
\usepackage[caption=false,font=normalsize,labelfont=sf,textfont=sf]{subfig}
\usepackage{textcomp}
\usepackage{stfloats}
\usepackage{url}
\usepackage{verbatim}
\usepackage{graphicx}
\usepackage{cite}
\usepackage{booktabs} 
\usepackage{multirow}
\usepackage{xcolor}
\usepackage[super]{nth}
\usepackage{tikz}
\newcommand*\circled[1]{\tikz[baseline=(char.base)]{
            \node[shape=circle,draw,inner sep=0.2pt] (char) {#1};}}
\newcommand*\circledB[1]{\tikz[baseline=(char.base)]{
            \node[shape=circle,fill,inner sep=0.2pt] (char) {\textcolor{white}{#1}};}}
\usepackage{soul}
\usepackage{enumitem}

\begin{document}


\title{\huge PENDRAM: Enabling High-Performance and Energy-Efficient Processing of Deep Neural Networks through a Generalized DRAM Data Mapping Policy}

\author{Rachmad Vidya Wicaksana Putra,~\IEEEmembership{Member,~IEEE,} Muhammad Abdullah Hanif,~\IEEEmembership{Member,~IEEE,} and \\ Muhammad Shafique,~\IEEEmembership{Senior Member,~IEEE} 
\thanks{Rachmad Vidya Wicaksana Putra and Muhammad Abdullah Hanif are with eBrain Lab, Division of Engineering, New York University (NYU) Abu Dhabi, United Arab Emirates;
{e-mail: \{rachmad.putra, mh6117\}@nyu.edu}}
\thanks{Muhammad Shafique is the Director of eBrain Lab, Division of Engineering, New York University (NYU) Abu Dhabi, United Arab Emirates;
{e-mail: muhammad.shafique@nyu.edu}}
\thanks{Manuscript received Month DD, 2024; revised Month DD, 2024.}
\vspace{-1cm}
}

\markboth{Journal of \LaTeX\ Class Files,~Vol.~XX, No.~X, Month~YYYY}%
{Shell \MakeLowercase{\textit{et al.}}: A Sample Article Using IEEEtran.cls for IEEE Journals}


\maketitle


\begin{abstract}
Convolutional Neural Networks (CNNs), a prominent type of Deep Neural Networks (DNNs), have emerged as a state-of-the-art solution for solving machine learning tasks. 
To improve the performance and energy efficiency of CNN inference, the employment of specialized hardware accelerators is prevalent.
However, CNN accelerators still face performance- and energy-efficiency challenges due to high off-chip memory (DRAM) access latency and energy, which are especially crucial for latency- and energy-constrained embedded applications. 
Moreover, different DRAM architectures have different profiles of access latency and energy, thus making it challenging to optimize them for high performance and energy-efficient CNN accelerators.
To address this, we present PENDRAM, a novel design space exploration methodology that enables high-performance and energy-efficient CNN acceleration through a generalized DRAM data mapping policy. 
Specifically, it explores the impact of different DRAM data mapping policies and DRAM architectures across different CNN partitioning and scheduling schemes on the DRAM access latency and energy, then identifies the pareto-optimal design choices.  
The experimental results show that our DRAM data mapping policy improves the energy-delay-product of DRAM accesses in the CNN accelerator over other mapping policies by up to 96\%. 
In this manner, our PENDRAM methodology offers high-performance and energy-efficient CNN acceleration under any given DRAM architectures for diverse embedded AI applications.
\end{abstract}


\vspace{-0.3cm}
\section{Introduction}
\label{Sec_Intro}

\begin{figure}[t]
  \centering
  \includegraphics[width=0.95\linewidth]{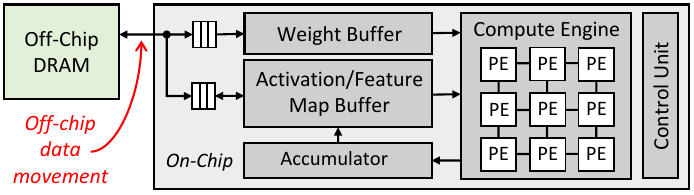}
  \vspace{-0.2cm}
  \caption{The typical HW architecture of CNN accelerators.
  Here, each processing element (PE) represents a Multiply-and-Accumulate (MAC) operation.}
  \label{Fig_CNN_Accel}
  \vspace{-0.5cm}
\end{figure}

\IEEEPARstart{I}{n} recent years, the widespread use of Convolutional Neural Networks (CNNs), a prominent type of Deep Neural Networks (DNNs), for organizing, analyzing, and inferring information from digital data is growing fast. 
The reason is that, CNNs have achieved state-of-the-art performance in solving a wide range of Machine Learning (ML) applications, such as image classification, object segmentation, autonomous driving, smart health-care assistance, and even financial analysis~\cite{Ref_LeCun_DeepLearning_Nature15}.
Since the CNN algorithms are data- and compute-intensive, many specialized hardware (HW) accelerators have been proposed in the literature over the past few years to expedite the inference process~\cite{Ref_Chen_DianNao_ASPLOS14, Ref_Zhang_CNNfpga_FPGA15, Ref_Zhang_CambriconX_MICRO15, Ref_Han_EIE_ISCA16, Ref_Chen_Eyeriss_JSSC16,  Ref_Albericio_Cnvlutin_ISCA16, Ref_Luo_DaDianNao_TC17, Ref_Jouppi_TPU_ISCA17, Ref_Parashar_SCNN_ISCA17, Ref_Lu_FlexFlow_HPCA17, Ref_Moons_Envision_ISSCC17,  Ref_Yuan_Sticker_VLSIC18, Ref_Kwon_MAERI_ASPLOS18, Ref_Sharma_BitFusion_ISCA18, Ref_Ueyoshi_QUEST_ISSCC18, Ref_Zhang_SNAP_VLSIC19, Ref_Lee_UNPU_JSSC19, Ref_Sayal_CNNengine_JSSC20, Ref_Putra_MPNA_Springer23}.
These specialized HW accelerators can achieve significantly higher performance and energy efficiency as compared to general-purpose CPUs~\cite{Ref_Shafique_TowardsEdgeAI_ICCAD21}; see their typical hardware achitecture in Fig.~\ref{Fig_CNN_Accel}. 
However, they still face performance- and energy-efficiency challenges due to the high DRAM-based off-chip memory access latency and energy, which are significantly higher than the latency and energy for the other compute operations~\cite{Ref_Sze_DNNsurvey_IEEE17}\cite{Ref_Putra_ROMANet_TVLSI21}. 
Therefore, \textit{reducing the DRAM access latency and energy is the key for improving the performance- and energy-efficiency of CNN accelerators}.
Moreover, we observe that different DRAM architectures have different characteristics regarding their access energy and access latency~\cite{Ref_Kim_SALP_SIGARCH12}\cite{Ref_Lee_TLDRAM_HPCA13}. 
For instance, new DRAM architectures, such as the DRAM with Subarray-Level Parallelism (SALP)~\cite{Ref_Kim_SALP_SIGARCH12} and the Tiered-Latency DRAM (TL-DRAM)~\cite{Ref_Lee_TLDRAM_HPCA13}, have been proposed to address limitations of commodity DRAMs in terms of latency with some overheads in area and power/energy consumption, thereby having the potential for improving the efficiency of CNN accelerators.
Therefore, \textit{an efficient technique is required to benefit from different DRAM architectures for reducing the DRAM access latency and energy with minimum overheads}.

\vspace{-0.2cm}
\subsection{The State-of-the-Art and Their Limitations}
\label{Sec_Intro_SoA}

The existing works have proposed different techniques to reduce DRAM access energy by focusing on minimizing the number of DRAM accesses~\cite{Ref_Zhang_CNNfpga_FPGA15}\cite{Ref_Putra_ROMANet_TVLSI21}\cite{Ref_Li_SmartShuttle_DATE18, Ref_Stoutchinin_Scheduling_arXiv19, Ref_Tewari_BuswidthOffChip_ISVLSI20}. 
They have similar key ideas: 
(1) defining data partitioning\footnote{\scriptsize Data partitioning determines portions of data in the form of block/tile to be accessed from off-chip DRAM to on-chip memory/buffer at one time. A detailed explanation is provided in Section \ref{Sec_Prelim_PartitionSchedule}.} and then transferring each partition from the off-chip DRAM to the on-chip memory/buffer in a defined schedule, and 
(2) maximally reusing the data that are already in the on-chip memory/buffer.
Furthermore, the state-of-the-art works \cite{Ref_Li_SmartShuttle_DATE18}\cite{Ref_Putra_ROMANet_TVLSI21} also consider adaptive data partitioning and scheduling to minimize the number of DRAM accesses, i.e., by adaptively switching the reuse priority among different data types: input activations/feature maps (\textit{ifms}), output activations/feature maps (\textit{ofms}), and filter weights (\textit{wghs}), across the layers of a network. 
Although these state-of-the-art works result in a reduced number of DRAM accesses (that also leads to a reduced DRAM access energy), \textit{they have not studied the characteristics and potentials of different DRAM architectures for CNN accelerators, thereby limiting their performance and energy efficiency gains}.

\vspace{-0.3cm}
\subsection{Motivational Case Study and Scientific Challenges}
\label{Sec_Intro_Challenges}

\textbf{Overview:}
Although there are different types of commodity DRAMs (e.g., DDR3 and DDR4), they employ similar internal organization and operations~\cite{Ref_Ghose_DRAMworkload_POMACS19}, thereby having similar trends of latency-per-access and energy-per-access across different access conditions (such as row buffer hit, miss, and conflict)\footnote{The details of DRAM fundamentals are discussed in Section~\ref{Sec_Prelim_DRAMorg}}. 
In commodity DRAMs, each request is directed to a bank and it can only access a subarray at a time, despite there are multiple subarrays-per-bank~\cite{Ref_Kim_SALP_SIGARCH12}.
This limits the DRAMs' capability to offer lower access latency and energy. 
To address this, new DRAM architectures have been proposed in the literature. 
For instance, \textit{subarray-level parallelism (SALP)} in a bank is enabled through three variants of architectures, including SALP-1, SALP-2, and SALP-MASA\footnote{The details of SALP architectures are discussed in Section~\ref{Sec_Prelim_SALP}}~\cite{Ref_Kim_SALP_SIGARCH12}.
Another work proposed \textit{tiered-latency DRAM (TL-DRAM)}~\cite{Ref_Lee_TLDRAM_HPCA13} which splits the long bitline in each subarray into two shorter segments (i.e., near and far segments from the sense amplifier) using an isolation transistor\footnote{The details of TL-DRAM architecture are discussed in Section~\ref{Sec_Prelim_TLDRAM}}. 

\begin{figure*}[t]
  \centering
  \includegraphics[width=0.82\linewidth]{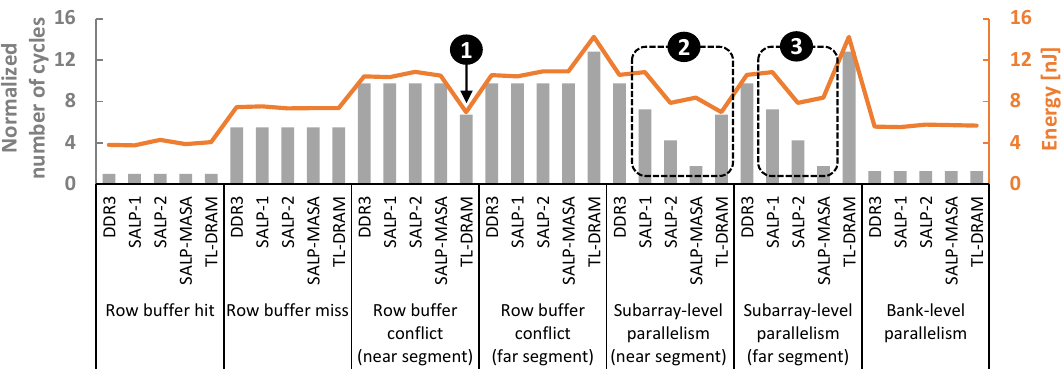}
  \vspace{-0.5cm}
  \caption{DRAM latency-per-access and energy-per-access for different access conditions (i.e., a row buffer hit, a row buffer miss, a row buffer conflict, subarray- and bank-level parallelism) in different DRAM architectures (DDR3, SALP-1, SALP-2, SALP-MASA, and TL-DRAM). Data are obtained from our experiments using state-of-the-art cycle-accurate DRAM simulators~\cite{Ref_Kim_Ramulator_LCA15, Ref_Ghose_VAMPIRE_POMACS18} for DDR3-1600 2Gb x8 and SALP 2Gb x8 with 8 subarrays-per-bank.}
  \label{Fig_HitMissConflict}
  \vspace{-0.5cm}
\end{figure*}

\textbf{Case Study:}
To understand the characteristics of different DRAM architectures, we perform an experimental case study\footnote{The details of experimental setup are discussed in Section~\ref{Sec_EvalMethod}} to observe the DRAM latency-per-access and energy-per-access considering different access conditions, and the results are presented in Figure~\ref{Fig_HitMissConflict}.
Our observation results show that \textit{SALP and TL-DRAM architectures have the potential to further reduce the DRAM latency-per-access and energy-per-access as compared to commodity DRAMs}, since they can offer lower latency and/or energy consumption in certain conditions; as shown by \circledB{1}, \circledB{2}, and \circledB{3} in Figure~\ref{Fig_HitMissConflict}. 

\textbf{Challenges:}
Results of our case study show that the DRAM latency-per-access and energy-per-access can be further optimized considering the given DRAM architecture to improve the energy efficiency of DRAM accesses for CNN accelerators.
Hence, there is a need for \textit{a generalized DRAM mapping policy} that can achieve maximum row buffer hits while exploiting the internal organization of DRAM. 
Furthermore, to justify that our DRAM mapping policy is applicable to different design choices, \textit{a design space exploration (DSE)} is required. 
This DSE should investigate the impact of different DRAM mapping policies in different DRAM architectures with different data partitioning and scheduling schemes, to find the minimum energy-delay-product (EDP) of DRAM accesses. 
This EDP is a measure of the energy efficiency of a CNN accelerator.
To enable this, \textit{an analytical model to estimate the EDP of different DRAM mapping policies in DSE} is also needed. 

\textit{\textbf{Required:} A design methodology that leverages a generalized DRAM mapping policy to optimize DRAM energy-per-access and latency-per-access considering different DRAM architectures for CNN accelerators.}

\vspace{-0.2cm}
\subsection{Our Novel Contributions}
\label{Sec_Intro_Novelty}

\begin{figure}[t]
  \centering
  \includegraphics[width=\linewidth]{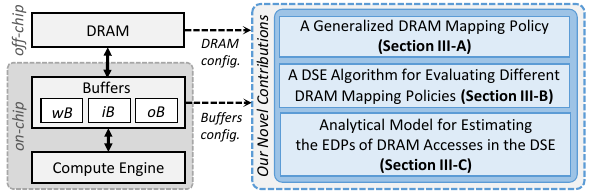}
  \vspace{-0.5cm}
  \caption{The overview of our novel contributions, highlighted in blue. In this work, we consider separate on-chip buffers in a CNN accelerator for different data types, i.e., input buffer (\textit{iB}) for \textit{ifms}, weight buffer (\textit{wB}) for \textit{wghs}, and output buffer (\textit{oB}) for \textit{ofms}.}
  \label{Fig_Novelty}
  \vspace{-0.5cm}
\end{figure}

To address the above challenges, we propose \textit{PENDRAM, a novel methodology to enable high-\underline{P}erformance and \underline{E}nergy-efficient CNN accelerators through a generalized \underline{DRAM} data mapping policy}. 
It employs the following key techniques; see an overview in Figure~\ref{Fig_Novelty}.
\begin{enumerate}[leftmargin=*]
  \item We propose \textbf{a generalized DRAM data mapping policy} that leads to minimum EDP of DRAM accesses, for any given combination of DRAM architecture, data partitioning, and scheduling scheme in a CNN accelerator.
  Its key idea is to orderly prioritizes maximizing row buffer hits, bank-level parallelism, and subarray-level parallelism in the near segment of the subarray.
  \item We propose \textbf{a DSE algorithm} to find a DRAM mapping policy that offers minimum EDP of DRAM accesses, while considering different DRAM architectures, different data partitioning, and scheduling schemes.
  \item We propose \textbf{an analytical model for estimating EDP of DRAM mapping policies in the DSE}. 
  Here, the EDP for each DRAM mapping policy is estimated by multiplying the number of DRAM accesses with the respective number of cycles and energy consumption. 
\end{enumerate}

\vspace{-0.2cm}
\section{Preliminaries}
\label{Sec_Prelim}

\subsection{Data Partitioning and Scheduling for CNN Processing}
\label{Sec_Prelim_PartitionSchedule}
\begin{figure}[hbtp]
  \centering
  \vspace{-0.2cm}
  \includegraphics[width=\linewidth]{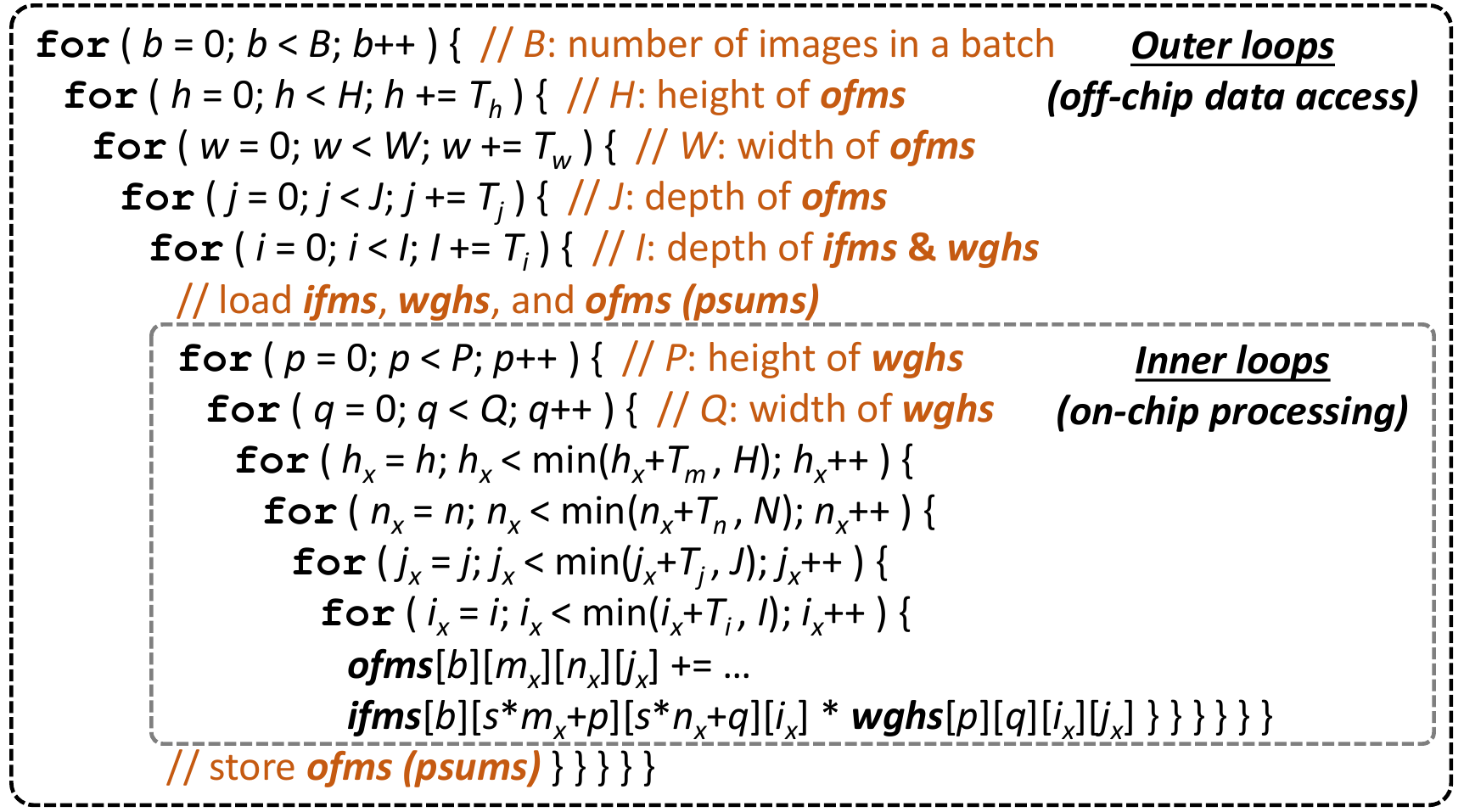}
  \vspace{-0.7cm}
  \caption{Pseudo-code of the tiled CNN processing. Here, inner loops represent the on-chip processing, outer loops represent the off-chip data access, and $s$ denotes the stride of convolution.}
  \label{Fig_PseudoCode_CNN}
  \vspace{-0.6cm}
\end{figure}

The full CNN model usually cannot be mapped at once on the HW accelerator fabric because of the limited capacity of on-chip buffers (i.e., about $100$KB-$500$KB~\cite{Ref_Sze_DNNsurvey_IEEE17}), hence data partitioning and scheduling are required to complete the CNN processing.
A pseudo-code of a convolutional processing in a CNN accelerator is illustrated in Fig.~\ref{Fig_PseudoCode_CNN}.
Here, the inner loops represent the on-chip processing. 
Meanwhile, the outer loops represent the scheduling of processing different partitions of data from all data types (i.e., \textit{ifms}, \textit{wghs}, and \textit{ofms}). 
These data are partitioned in the form of tiles, whose sizes have to be less than or equal to the sizes of respective buffers (i.e., \textit{iB}, \textit{wB}, and \textit{oB}). 
Furthermore, the sequence of the outer loops represents the order in which the tiles are accessed from DRAM to the on-chip buffer, thus reflecting the number of DRAM accesses required to process a layer of a network.

\subsection{DRAM Fundamentals}
\label{Sec_Prelim_DRAMorg}

\begin{figure}[t]
  \centering
  \includegraphics[width=\linewidth]{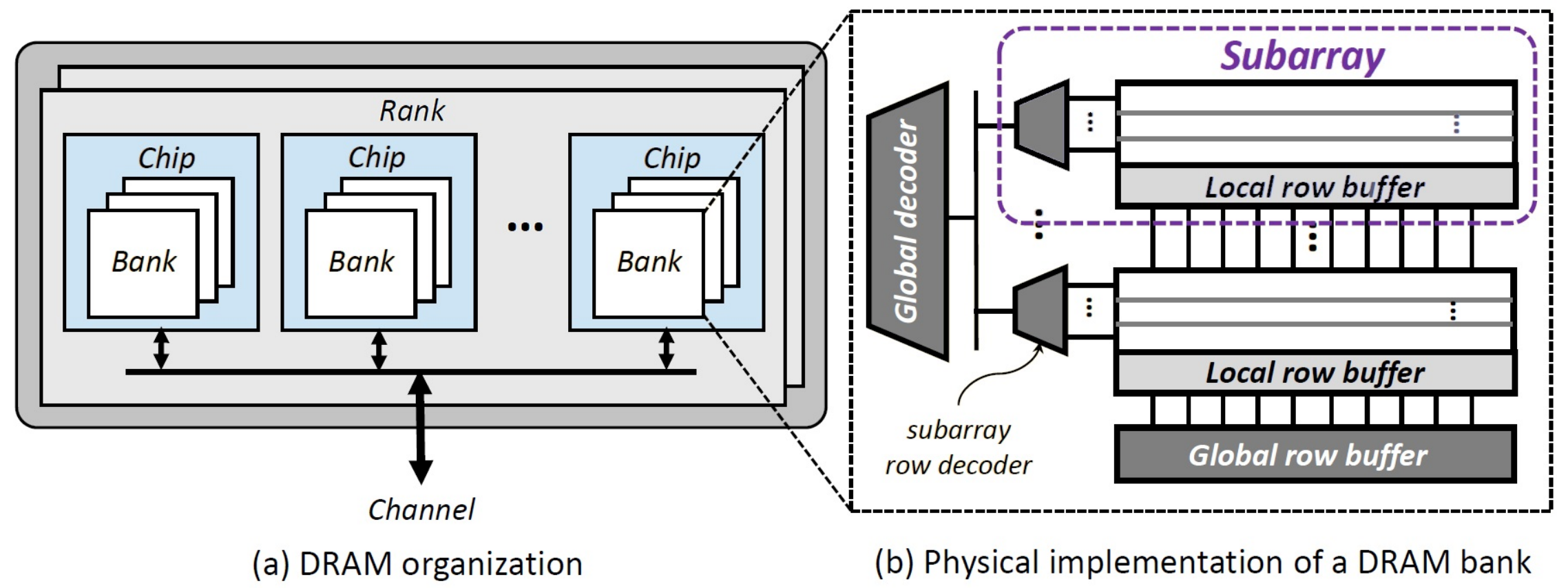}
  \vspace{-0.5cm}
  \caption{(a) Overview of the DRAM organization. (b) Physical implementation of a DRAM bank, showing multiple subarrays in a bank.}
  \label{Fig_DRAMorg}
  \vspace{-0.5cm}
\end{figure}

\textbf{Organization:} 
From top-to-down hierarchy, the organization of a commodity DRAM comprises \textit{channel}, \textit{rank}, \textit{chip}, \textit{bank}, \textit{row}, and \textit{column} \cite{Ref_Ghose_DRAMworkload_POMACS19}\cite{Ref_Liu_RAIDR_SIGARCH12}, as shown in Fig.~\ref{Fig_DRAMorg}(a). 
Here, banks are the lowest hierarchy, which can be accessed in parallel (referred to as \textit{bank-level parallelism}) \cite{Ref_Kim_MemScheduling_MM11}.
Physically, A DRAM bank is \textit{not} implemented in a monolithic design through a large array of cells with a single row buffer. 
In fact, a bank is implemented in multiple subarrays, and each having its local row buffer; see Fig \ref{Fig_DRAMorg}(b). 
Multiple subarrays in a bank share \textit{global bitlines}, which connect local row buffers to a global row buffer, and \textit{a global row address decoder} \cite{Ref_Kim_SALP_SIGARCH12}.

\textbf{Operations:} 
A specific rank will respond to each DRAM request, and multiple chips in this rank can be accessed in parallel, contributing to a DRAM word.
In each chip, the request is directed to a specific bank and decoded into row and column addresses.
Here, the \textit{Activation} (ACT) command triggers a row activation, and data from the requested row are copied to the row buffer.
Then, a \textit{read} (RD) or \textit{write} (WR) command can be executed to the requested column in the activated row buffer.
Here, a DRAM request may encounter different possible access conditions.
If the requested row is already activated, then it is \textit{a row buffer hit}.  
If the requested row is not activated, then it is either \textit{a row buffer miss} or \textit{conflict}. 
In \textit{a row buffer miss}, there is no activated row in the row buffer, hence this condition requires to activate the requested row.
Meanwhile, in \textit{a row buffer conflict}, there is an activated row in the row buffer, but it is \textit{not} the requested row. 
Hence, this condition requires to close the activated row using the \textit{precharging} (PRE) command, then activate the requested row using the \textit{activation} (ACT) command.

\textbf{Data Mapping:} 
The default DRAM data mapping policy stores the data to different columns of the same row for maximizing row buffer hits, and to different banks of the same rank for maximizing bank-level parallelism. 
However, it does not exploit subarray-level parallelism and does not consider data partitioning and scheduling schemes in CNN processing, hence the default data mapping solution is sub-optimal.

\vspace{-0.2cm}
\subsection{Subarray-Level Parallelism (SALP)-enabled DRAM}
\label{Sec_Prelim_SALP}

To reduce latency with minimum area and energy overheads, a recent work proposed three variants of DRAM architectures and mechanisms that exploit SALP in a DRAM bank, called SALP-1, SALP-2, and SALP-MASA~\cite{Ref_Kim_SALP_SIGARCH12}, whose key ideas are the following.
\begin{itemize}[leftmargin=*]
    \item \textbf{SALP-1} reduces the DRAM service time by overlapping the \textit{precharging} of one subarray with the \textit{activation} of another subarray. 
    To do this, re-interpretation of the existing timing constraint for \textit{precharging}, is needed.
    \item \textbf{SALP-2} reduces the DRAM service time by overlapping the \textit{write-recovery} latency of an active subarray, with the \textit{activation} of another subarray. 
    To do this, additional circuits to activate two subarrays at the same time is needed.
    \item \textbf{Multitude of Activated Subarrays (MASA)} reduces the DRAM service time by activating multiple subarrays at the same time. 
    To do this, additional circuits to activate multiple subarrays at the same time is needed.
\end{itemize}
For further details on the SALP-enabled DRAM architectures, we refer to the original studies in~\cite{Ref_Kim_SALP_SIGARCH12}.

\vspace{-0.2cm}
\subsection{Tiered-Latency DRAM (TL-DRAM)}
\label{Sec_Prelim_TLDRAM}

\begin{figure}[hbtp]
  \vspace{-0.4cm}
  \centering
  \includegraphics[width=\linewidth]{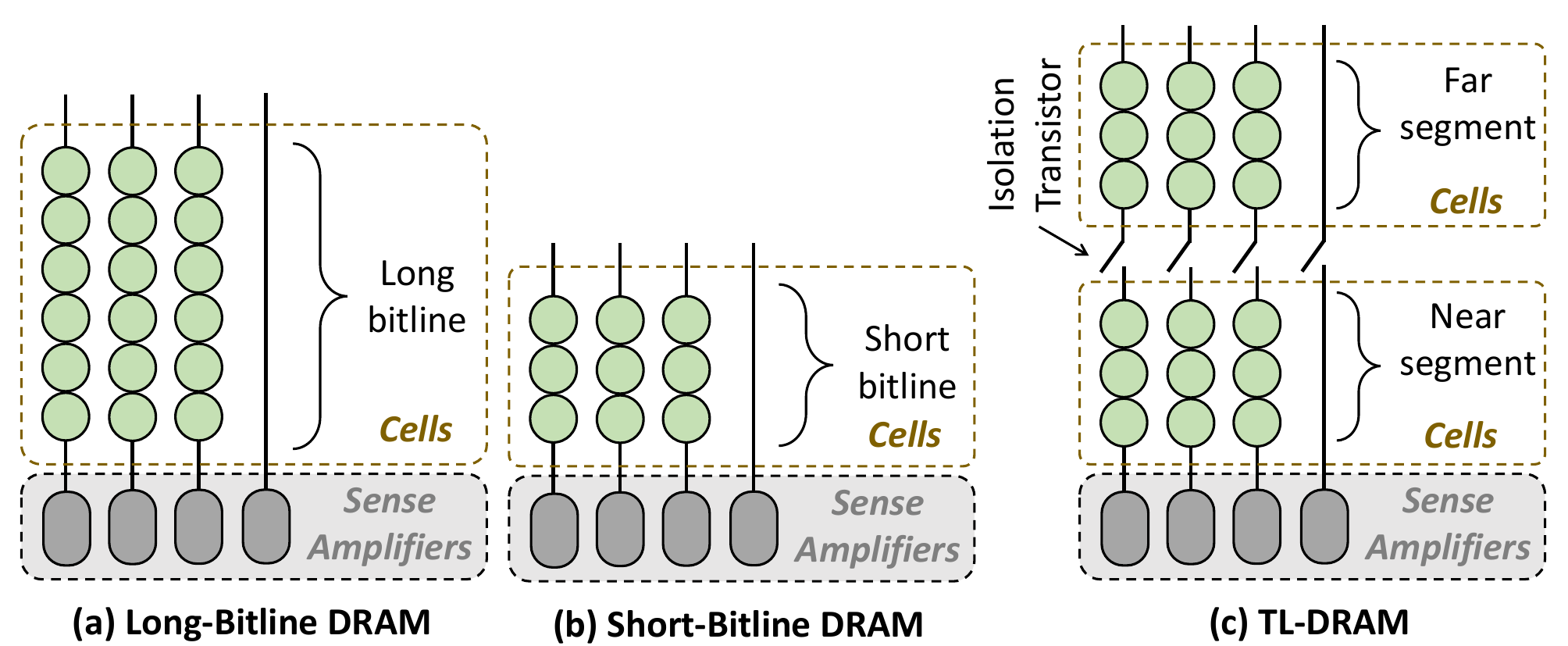}
  \vspace{-0.6cm}
  \caption{Bitlines in (a) long-bitline DRAM, (b) shart-bitline DRAM, and (c) TL-DRAM with near and far segments.}
  \label{Fig_TLDRAMorg}
  \vspace{-0.6cm}
\end{figure}

In commodity DRAMs, many storage cells are connected to a sense amplifier through a single bitline, for optimizing the DRAM \textit{cost-per-bit}~\cite{Ref_Lee_TLDRAM_HPCA13}. 
In this manner, many cells can be sensed with a relatively small number of sense amplifiers, but at the cost of long bitlines which lead to \textit{a high parasitic capacitance} (the dominant source of DRAM latency)~\cite{Ref_Lee_TLDRAM_HPCA13}. 
To achieve both low cost-per-bit and low access latency, a recent work proposed TL-DRAM architecture, which splits long bitline in each subarray into two shorter segments (i.e., \textit{near} and \textit{far segments}) using an isolation transistor~\cite{Ref_Lee_TLDRAM_HPCA13}; see Fig.~\ref{Fig_TLDRAMorg}.
Therefore, the near segment can be accessed with the latency of a short bitline, while incurring minimum area and cost-per-bit overheads.
For further details on the TL-DRAM architecture, we refer to the original studies in~\cite{Ref_Lee_TLDRAM_HPCA13}.

\section{The PENDRAM Methodology}
\label{Sec_PENDRAM}

\begin{figure}[hbtp]
\vspace{-0.3cm}
\centering
\includegraphics[width=\linewidth]{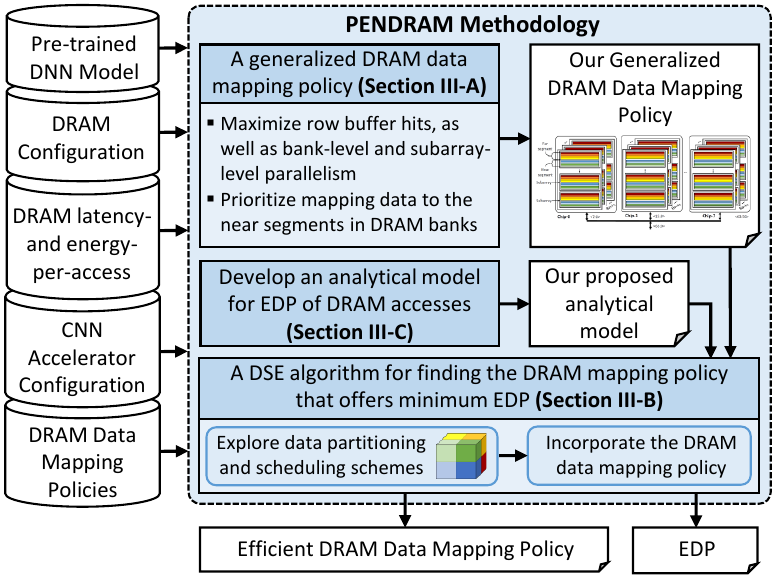}
\vspace{-0.5cm}
\caption{An overview of the PENDRAM methodology. The novel contributions are highlighted in blue.}
\label{Fig_PENDRAM}
\vspace{-0.6cm}
\end{figure}

We propose PENDRAM methodology to optimize DRAM energy-per-access and latency-per-access for a given DRAM architecture, data partitioning, and scheduling scheme in CNN accelerators. 
The key techniques in our PENDRAM are shown in Figure~\ref{Fig_PENDRAM}, and described in Section~\ref{Sec_PENDRAM_DRAMmapping} to Section~\ref{Sec_PENDRAM_Model}.

\subsection{The Generalized DRAM Data Mapping Policy}
\label{Sec_PENDRAM_DRAMmapping}

\begin{figure}[t]
  \centering
  \includegraphics[width=\linewidth]{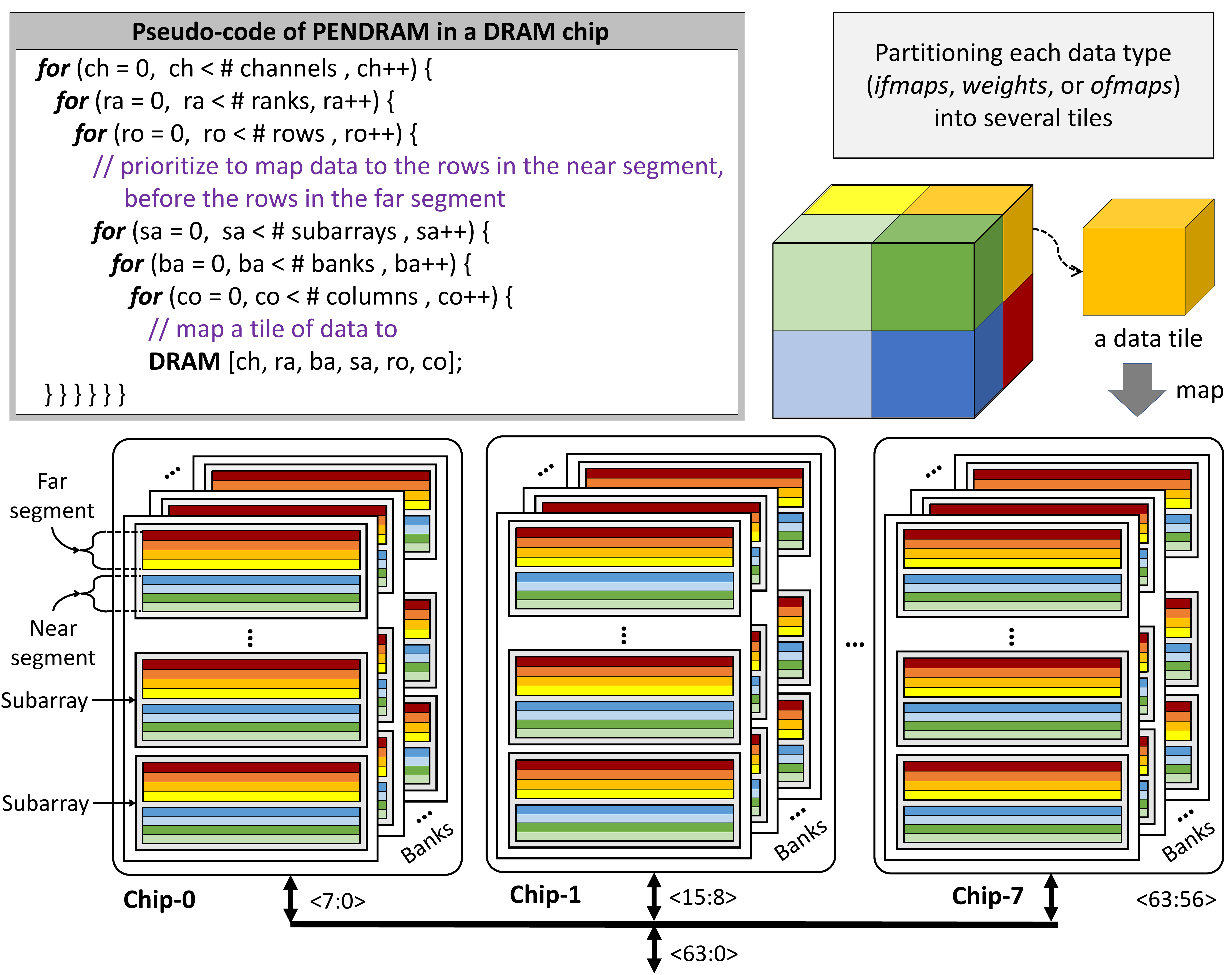}
  \vspace{-0.6cm}
  \caption{Pseudo-code of our DRAM data mapping policy and its conceptual implementation in DRAM.}
  \label{Fig_PENDRAM_MapPhysical}
  \vspace{-0.6cm}
\end{figure}

Results of our experimental case study in Fig.~\ref{Fig_HitMissConflict} show that different DRAM architectures have similar patterns in terms of latency-per-access and energy-per-access across different access conditions (e.g., row buffer hits, misses, and conflicts). 
These patterns can be exploited to achieve efficient DRAM access latency and energy.
Toward this, \textit{we propose a generalized DRAM data mapping policy to achieve efficient DRAM access latency and energy for any given DRAM architecture as well as data partitioning and scheduling scheme}, thereby enabling high-performance and energy-efficient CNN accelerators.
Its main idea is to orderly prioritize the data mapping that maximizes DRAM row buffer hit, bank-parallelism, and subarray-level parallelism in the near segments of the subarrays over the far segments.
The pseudo-code and physical representation of the proposed DRAM mapping policy are shown in Fig.~\ref{Fig_PENDRAM_MapPhysical}. 
Our DRAM data mapping policy considers tile-based partitioning, hence it can be performed for each data tile using the following steps (in each DRAM chip).
\smallskip
\\
\textbf{\underline{Step-1:}} 
    \begin{itemize}[leftmargin=*]
        \item We identify a target row that is available and closer to the sense amplifier in the target subarray and target bank, thus \textit{prioritizing rows in the near segment over the far segment}. 
        \item Then, we map the data to different columns in the target row \textit{to maximize row buffer hits}.
        \item If multiple chips are available within a rank, this step can be performed in different chips in parallel \textit{for exploiting the chip-level parallelism}.
        \item If some data remain but the target row of the target subarray and the target bank is full, then we go to \textbf{Step-2}.
    \end{itemize}
\textbf{\underline{Step-2:}} 
    \begin{itemize}[leftmargin=*]
        \item We map the remaining data to a different target bank in the same chip \textit{to exploit bank-level parallelism}.
        \item If multiple chips are available, then this step can be performed in different chips in parallel.
        \item Then, we follow the mapping mechanism in \textbf{Step-1} to \textbf{Step-2} again until all data are mapped in the target subarray across all banks. 
        \item If some data remain but the target row and target subarray across all banks are full, then we go to \textbf{Step-3}. 
    \end{itemize}
\textbf{\underline{Step-3:}}
    \begin{itemize}[leftmargin=*]
        \item We map the remaining data to a different target subarray in the target bank \textit{to exploit subarray-level parallelism}.
        \item If multiple chips are available, then this step can be performed in different chips in parallel. 
        \item Then, we follow \textbf{Step-1} to \textbf{Step-3} again until all data are mapped in the target subarray across all banks. 
        \item If some data remain but the target row across all subarray and all banks are full, then we go to \textbf{Step-4}. 
    \end{itemize}
\textbf{\underline{Step-4:}} 
    \begin{itemize}[leftmargin=*]
        \item We select a different row index that is available and closer to sense amplifier from the target subarray and the target bank, as the new target row. 
        \item Then, we follow \textbf{Step-1} to \textbf{Step-4} until all data are mapped. 
        \item If some data remain but all memory cells in a rank are full, then we perform \textbf{Step-1} to \textbf{Step-4} again until all data are mapped to a different rank and channel, respectively. 
    \end{itemize}

\smallskip
To illustrate that our DRAM data mapping policy always achieves the minimum EDP of DRAM accesses in different possible settings, we perform an extensive DSE. 
Our DSE investigates different combinations of DRAM data mapping policies, different DRAM architectures (e.g., DDR3, SALP-1, SALP-2, SALP-MASA, and TL-DRAM), and different data partitioning and scheduling schemes, then estimates the EDP for these different combinations. 
\textit{This DSE is important to corroborate that the best solution that provides the minimum EDP in each given combination is always the same as provided by our DRAM data mapping policy.}

\vspace{-0.2cm}
\subsection{DSE for Evaluating Different DRAM Mapping Policies}
\label{Sec_PENDRAM_DSE}

\begin{figure}[hbtp]
\vspace{-0.2cm}
\centering
\includegraphics[width=\linewidth]{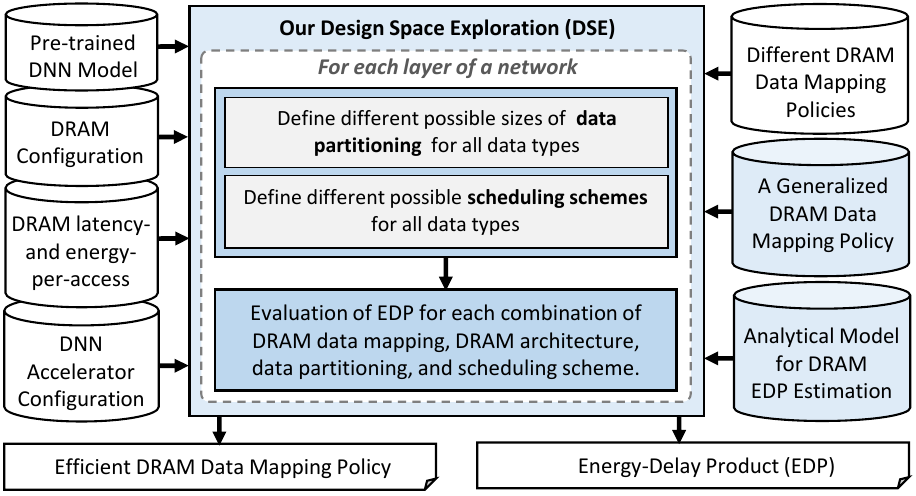}
\vspace{-0.6cm}
\caption{The operational flow of our DSE, showing the key steps.}
\label{Fig_PENDRAM_DSE}
\vspace{-0.1cm}
\end{figure}

To evaluate the impact of different DRAM data mapping policies and the performance of our proposed mapping policy as compared to the others, we perform an extensive DSE. 
An overview of the DSE is shown in Fig.~\ref{Fig_PENDRAM_DSE} and its pseudo-code algorithm is presented in Alg.~\ref{Alg_PENDRAM_DSE}. 

\begin{algorithm}[t]
\caption{Pseudo-code of the proposed DSE algorithm}
\label{Alg_PENDRAM_DSE}
\footnotesize
\begin{algorithmic}[1]
\renewcommand{\algorithmicrequire}{\textbf{INPUT:}}
\renewcommand{\algorithmicensure}{\textbf{OUTPUT:}}
\REQUIRE \textbf{(1)} CNN configuration: number of layers ($L$);\\ 
  \textbf{(2)} Buffer size: for \textit{ifms} ($iB$), for \textit{wghs} ($wB$), for \textit{ofms} ($oB$); \\
  \textbf{(3)} Analytical models of EDP ($EDP$); \\ 
  \textbf{(4)} Data partitioning for \textit{ifms}, \textit{wghs}, and \textit{ofms} ($Partition$); \\ 
  \textbf{(5)} DRAM access scheduling ($Schedule$); \\ 
  \textbf{(6)} DRAM data mapping policies ($DRAMmaps$); \\ 
\ENSURE \textbf{(1)} Efficient DRAM data mapping policy ($map$); \\
  \textbf{(2)} Minimum EDP ($minEDP$); \\
\smallskip
\textbf{BEGIN} \\
  \textbf{Initialization}: \\
  \STATE $T_p = P$; // height of \textit{wghs} tile \\
  \STATE $T_q = Q$; // width of \textit{wghs} tile\\
  \STATE $EDP[] = 0$; \\
  \STATE $minEDP[] = 0$; \\
  \textbf{Process}: \\
  \FOR{($l = 1$ to $L$)}
    \FOR{(each $Partition$)}
      \FOR{(each $Schedule$)}
        \FOR{(each $DRAMmaps$)}
          \IF{(\textit{ifms} tile $\leq iB$) \AND (\textit{wghs} tile $\leq wB$) \AND (\textit{ofms} tile $\leq oB$)}
            \STATE Calculate $EDP[l]$;\\
            \IF{(first loop)}
              \STATE $minEDP[l] = EDP[l]$;\\
	      \ELSIF{($EDP[l] \leq minEDP[l]$)}
	       \STATE $minEDP[l] = EDP[l]$;\\
		   \STATE Save $map$, $minEDP$;\\
	      \ENDIF
	   \ENDIF
	\ENDFOR
      \ENDFOR
    \ENDFOR
\ENDFOR
\RETURN (1) $map$; (2) $minEDP$; \\
\textbf{END}
\end{algorithmic} 
\end{algorithm}
\setlength{\textfloatsep}{8pt}

For each layer of a network, our DSE employs three key steps: (1) defining different sizes of data tiles and scheduling schemes, (2) defining different DRAM data mapping policies, and (3) performing exploration to find a mapping policy that offers minimum EDP. 
The operational flow is explained in the following points.
\begin{itemize}[leftmargin=*]
    \item We define different sizes of data tiles for all data types (\textit{ifms}, \textit{wghs}, and \textit{ofms}), and different scheduling schemes. 
    \begin{itemize}
        \item Tile sizes are defined by the step sizes in the outer loops of CNN processing in Fig.~\ref{Fig_PseudoCode_CNN}.
        The tile sizes of the \textit{ifms}, \textit{wghs}, and \textit{ofms} have to fit in the corresponding buffers (\textit{iB}, \textit{wB}, and \textit{oB}).
        Each combination of the tile sizes for all data types defines one possible partitioning, which will be considered in the DSE.
        \item The scheduling schemes are defined by the sequence of the outer loops of CNN processing in Fig.~\ref{Fig_PseudoCode_CNN}. 
        Here, we consider four scheduling schemes, based on the reuse priority: 
        \textit{ifms-}, \textit{wghs-}, \textit{ofms-}, and \textit{adaptive-reuse} scheduling. 
        \begin{itemize}
            \item[\checkmark] The \textit{ifms-reuse} scheduling means that \textit{ifms} data type will be maximally reused when the data are available in the on-chip buffer. 
            A similar definition is applied for \textit{wghs-reuse} and \textit{ofms-reuse}.
            \item[\checkmark] The \textit{adaptive-reuse} scheduling means that the reuse priority changes across different layers of a network, according to which one among \textit{ifms}-/\textit{wghs}-/\textit{ofms-reuse} that offers the minimum number of DRAM accesses.
        \end{itemize}
    \end{itemize}
    \item We define different DRAM data mapping policies, by determining the different orders of mapping loops to different columns, rows (including near and far segments), subarrays, and banks in the same DRAM chip. 
    \begin{itemize}
        \item For commodity DRAMs, orders of mapping loops are the permutation of banks, rows, and columns. 
        \item For SALP-enabled DRAMs, orders of mapping loops are the permutation of banks, subarrays, rows, and columns.
        \item For TL-DRAM, orders of mapping loops are the permutation of banks, subarrays, (near and far segment) rows, and columns. 
    \end{itemize}
    \item We narrow down the design space by selecting the DRAM mapping policies that have the least frequent accesses to different rows, since it is the most expensive access in the same DRAM chip, for both latency and energy (as validated by Fig.~\ref{Fig_HitMissConflict}).
    Therefore, there are six mapping policies to be explored in the DSE, as presented in Table~\ref{Table_DRAMmaps}. 
    Note, our DRAM data mapping policy is represented as \textit{Mapping-3}.
    \item We perform the DSE to find a DRAM data mapping policy that offers minimum EDP, across different combinations of DRAM architectures, data partitioning, and scheduling schemes.
    The outputs of DSE are the minimum EDP and the DRAM corresponding mapping policy. 
\end{itemize}

\begin{table}[t]
  \caption{Different DRAM mapping policies for the DSE. Note, our proposed DRAM mapping policy is represented as \textit{Mapping-3}}
  \label{Table_DRAMmaps}
  \vspace{-0.2cm}
  \centering
  \footnotesize
  \begin{tabular}{|c|c|}
  \hline 
  \textbf{Mapping} & \textbf{Inner-most loops to outer-most loops} \\
  \hline
  \hline 
  1 & column, subarray, bank, row (near segment to far segment) \\
  \hline
  2 & subarray, column, bank, row (near segment to far segment) \\
  \hline
  3 & column, bank, subarray, row (near segment to far segment) \\
  \hline
  4 & bank, column, subarray, row (near segment to far segment) \\
  \hline
  5 & subarray, bank, column, row (near segment to far segment) \\
  \hline
  6 & bank, subarray, column, row (near segment to far segment) \\
  \hline
  \end{tabular}
  \vspace{-0.1cm}
\end{table}

Note, the DSE incorporates the characteristics of DRAM latency and energy for determining the EDP in the final results. 
For each layer of a network, the EDP value is obtained by multiplying the DRAM access latency and energy consumed by each combination of DRAM mapping policy, DRAM architecture, as well as configuration of data partitioning and scheduling scheme. 
Therefore, the DSE will be able to find the combination that offers minimum EDP for each layer of a network and minimum total EDP for a whole network. 

\vspace{-0.1cm}
\subsection{Analytical Model for EDP Estimation of DRAM Accesses}
\label{Sec_PENDRAM_Model}

Based on our DSE, \textit{the optimization problem is formulated to minimize the EDP of DRAM accesses for each layer of a network} and can be stated as the following.
\begin{equation}
  \footnotesize
  \begin{split}
    Objective:& \, minimize \, (EDP_{l})
  \label{Eq_OptimProblemLayer}
  \end{split}
\end{equation}
The EDP-per-layer ($EDP_{l}$) is obtained by multiplying the latency-per-layer and energy-per-layer. 
The latency-per-layer is obtained by accumulating all latency values incurred from the DRAM accesses for all data tiles during the processing of a network layer. 
Meanwhile, the energy-per-layer is obtained by accumulating all energy consumption values incurred from the DRAM accesses for all data tiles during the processing of a network layer.
Note, the DRAM access latency and energy are calculated on the basis of accesses for each data tile as we consider the tile-based data partitioning approach. 
For each data tile, the number of cycles required for DRAM accesses ($C_{tile}$) represents the DRAM access latency and can be calculated using Eq.~\ref{Eq_Ncycle_1tile}, while the DRAM access energy ($E_{tile}$) can be calculated using Eq.~\ref{Eq_Energy_1tile}. 
\begin{equation}
  \footnotesize
  \begin{split}
    C_{tile} = & acc_{column} \cdot C_{column} + acc_{row\_near} \cdot C_{row\_near} + \\
    & acc_{row\_far} \cdot C_{row\_far} + acc_{subarray} \cdot C_{subarray} + \\
    & acc_{bank} \cdot C_{bank}
    \label{Eq_Ncycle_1tile}
  \end{split}
\end{equation}
\begin{equation}
  \footnotesize
  \begin{split}
   E_{tile} = & acc_{column} \cdot E_{column} + acc_{row\_near} \cdot E_{row\_near} + \\
   & acc_{row\_far} \cdot E_{row\_far} + acc_{subarray} \cdot E_{subarray} + \\
   & acc_{bank} \cdot E_{bank}
   \label{Eq_Energy_1tile}
   \end{split}
\end{equation}
Here, term $acc_x$ represents the number of accesses to a different DRAM-$x$. 
$C_x$ represents the number of cycles incurred when accessing a different DRAM-$x$. 
$E_x$ represents the access energy incurred when accessing a different DRAM-$x$.
For all terms, $x \in$ \{\textit{column}, \textit{row\_near}, \textit{row\_far}, \textit{subarray}, \textit{bank}\}.
Note, \textit{row\_near} denotes row in the near segment, and \textit{row\_far} denotes row in the far segment.

\vspace{-0.1cm}
\section{Evaluation Methodology}
\label{Sec_EvalMethod}

\begin{figure}[hbtp]
\vspace{-0.1cm}
\centering
\includegraphics[width=\linewidth]{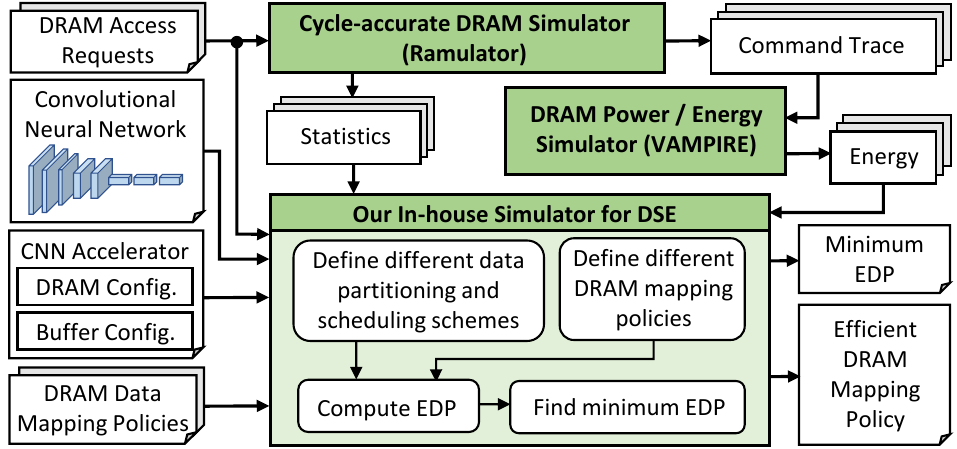}
\vspace{-0.6cm}
\caption{Experimental setup and tools flow.}
\label{Fig_ExpSetup}
\vspace{-0.2cm}
\end{figure}

To evaluate our proposed PENDRAM methodology, we built an experimental setup as shown in Fig.~\ref{Fig_ExpSetup}.
We employ a state-of-the-art cycle-accurate DRAM simulator, Ramulator~\cite{Ref_Kim_Ramulator_LCA15}, to obtain the statistics of latency (i.e, DRAM cycle-per-access) for different DRAM access conditions (e.g., row buffer hits, misses, conflicts, as well as subarray-level and bank-level parallelism) in different DRAM architectures.
Meanwhile, to profile the DRAM energy-per-access, we employ a real experiments-based DRAM energy simulator, VAMPIRE~\cite{Ref_Ghose_VAMPIRE_POMACS18}.
Information of the DRAM energy-per-access and cycle-per-access are then leveraged for the DSE, which considers different DRAM mapping policies, different DRAM architectures, as well as different data partitioning and scheduling schemes to find the DRAM mapping policy that offers minimum EDP.

In the DSE process, we consider a state-of-the-art TPU~\cite{Ref_Jouppi_TPU_ISCA17}-like CNN accelerator, as illustrated in Fig.~\ref{Fig_CNN_Accel}, with a reduced size of on-chip buffers and MAC array engine, as specified in Table~\ref{Table_DNNacc}.
We consider separate on-chip buffers for different data types (\textit{iB} for \textit{ifms}, \textit{wB} for \textit{wghs}, and \textit{oB} for \textit{ofms}).
To represent different DRAM architectures, we consider DDR3, SALPs (SALP-1, SALP-2, and SALP-MASA), as well as TL-DRAM. 
For scheduling schemes, we use \textit{ifms-reuse}, \textit{wghs-reuse}, \textit{ofms-reuse}, and \textit{adaptive-reuse} scheduling schemes.
For DRAM mapping policies, we evaluate six mapping policies presented in Table \ref{Table_DRAMmaps}.
For the inputs, we consider AlexNet~\cite{Ref_Alex_AlexNet_NIPS12}, VGG-16~\cite{Ref_Simonyan_VGG16_arXiv14}, MobileNet~\cite{Ref_Howard_MobileNet_arXiv17}, and SqueezeNet~\cite{Ref_Iandola_SqueezeNet_arXiv16} for dense networks, as well as a the Sparse MobileNet (that is obtained through the AutoML for Model Compression~\cite{Ref_He_AMC_ECCV18} technique), while considering the ImageNet dataset. 
\vspace{-0.1cm}
\begin{table}[hbtp]
\vspace{-0.2cm}
\caption{Configuration of the CNN accelerator.}
\label{Table_DNNacc}
\vspace{-0.2cm}
\centering
\footnotesize
\begin{tabular}{|l|l|}
\hline 
\multicolumn{1}{|c|}{\textbf{Module}} & \multicolumn{1}{c|}{\textbf{Description}} \\
\hline
\hline
CNN Processing Array & Size = 8$\times$8 MACs \\
\hline
On-chip Buffers & \textit{iB}: 64KB, \textit{wB}: 64KB, \textit{oB}: 64KB \\
\hline
\multirow{1}{*}{Memory Controller} &  Policy = open row, scheduler = FCFS\\
\hline
\multirow{3}{*}{DDR3-1600} & Configuration: 2Gb x8\\ 
     & 1 channel, 1 rank-per-channel \\
     & 1 chip-per-rank, 8 banks-per-chip \\
\hline
\multirow{4}{*}{SALPs} & Configuration: 2Gb x8\\ 
     & 1 channel, 1 rank-per-channel, \\
     & 1 chip-per-rank, 8 banks-per-chip, \\
     & 8 subarrays-per-bank \\
\hline
\multirow{5}{*}{TL-DRAM} & Configuration: 2Gb x8\\ 
     & 1 channel, 1 rank-per-channel, \\
     & 1 chip-per-rank, 8 banks-per-chip, \\
     & 32 subarrays-per-bank, \\
     & 64 rows/near segment, 960 rows/far segment \\
\hline
\end{tabular}
\vspace{-0.2cm}
\end{table}

\section{Results and Discussion}
\label{Sec_Results}

\begin{figure*}[t]
\centering
\includegraphics[width=0.8\linewidth]{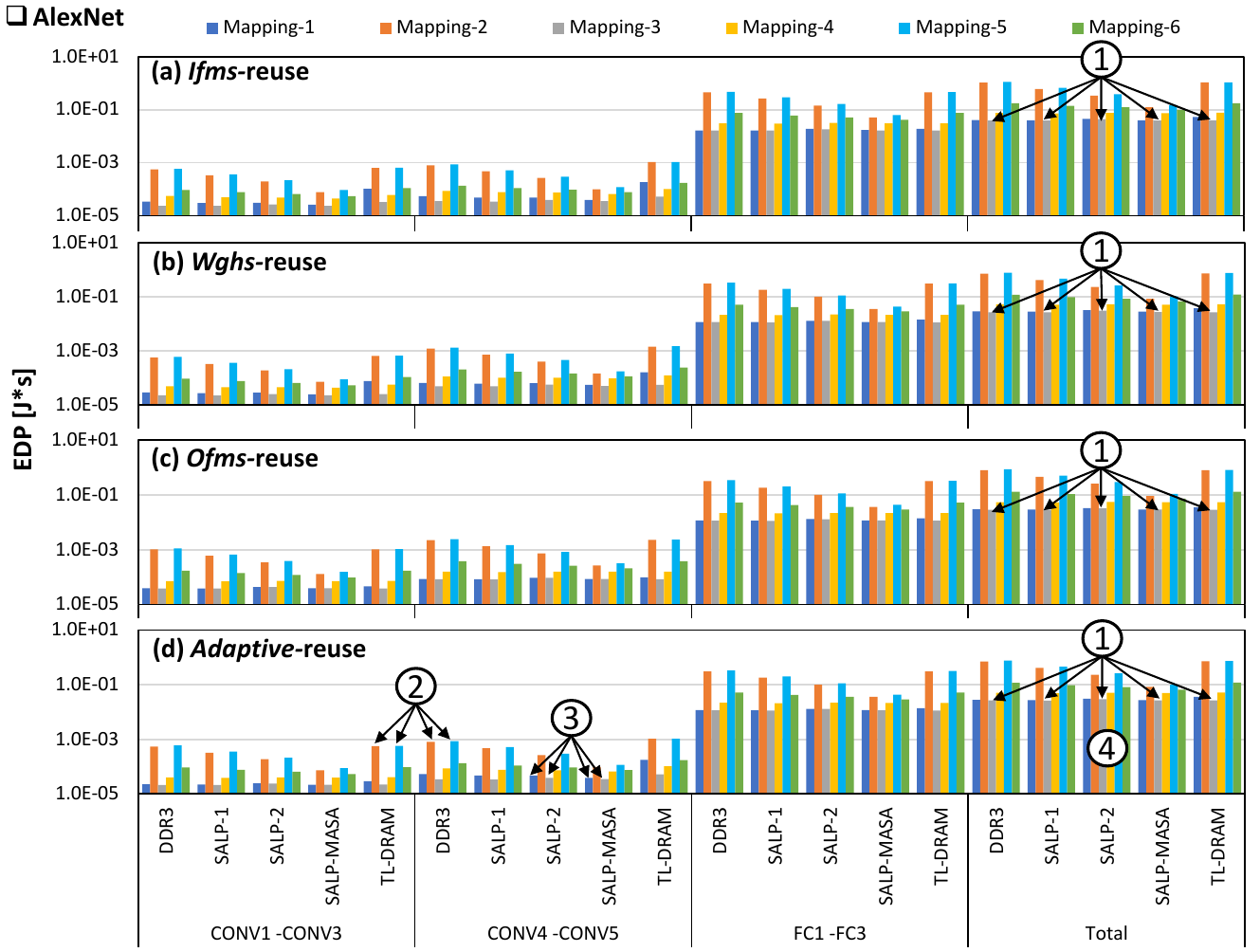}
\vspace{-0.3cm}
\caption{EDP of different DRAM mapping policies for AlexNet across different DRAM architectures (i.e., DDR3, SALP-1, SALP-2, SALP-MASA, and TL-DRAM), as well as different data partitioning and scheduling schemes: (a) \textit{ifms}-reuse, (b) \textit{wghs}-reuse, (c) \textit{ofms}-reuse, and (d) adaptive-reuse.}
\label{Fig_Result_AlexNet}
\vspace{-0.5cm}
\end{figure*}

\begin{figure*}[t]
\centering
\includegraphics[width=0.8\linewidth]{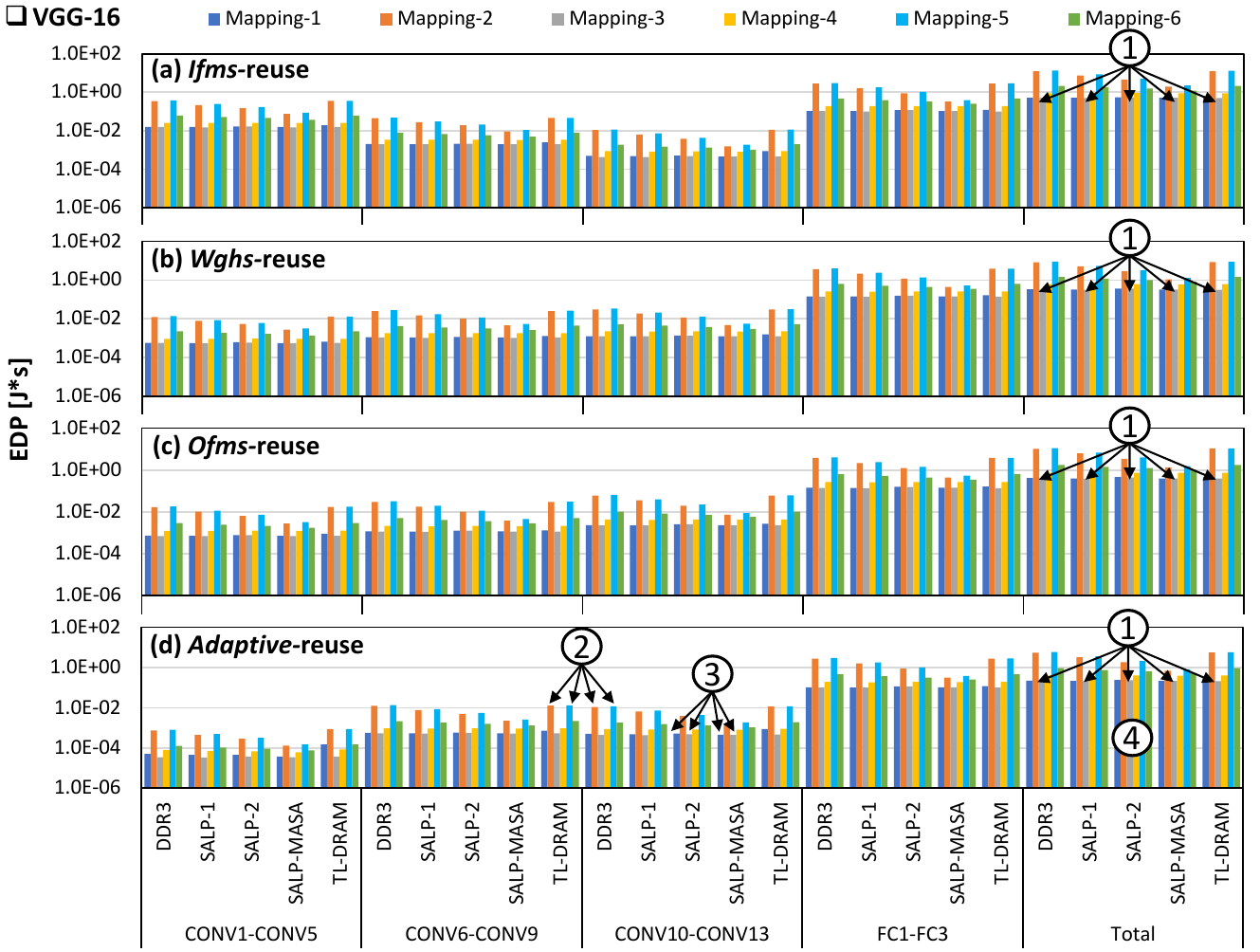}
\vspace{-0.2cm}
\caption{EDP of different DRAM mapping policies for VGG-16 across different DRAM architectures (i.e., DDR3, SALP-1, SALP-2, SALP-MASA, and TL-DRAM), as well as different data partitioning and scheduling schemes: (a) \textit{ifms}-reuse, (b) \textit{wghs}-reuse, (c) \textit{ofms}-reuse, and (d) adaptive-reuse.}
\label{Fig_Result_VGG16}
\vspace{-0.5cm}
\end{figure*}

\begin{figure*}[t]
\centering
\includegraphics[width=0.81\linewidth]{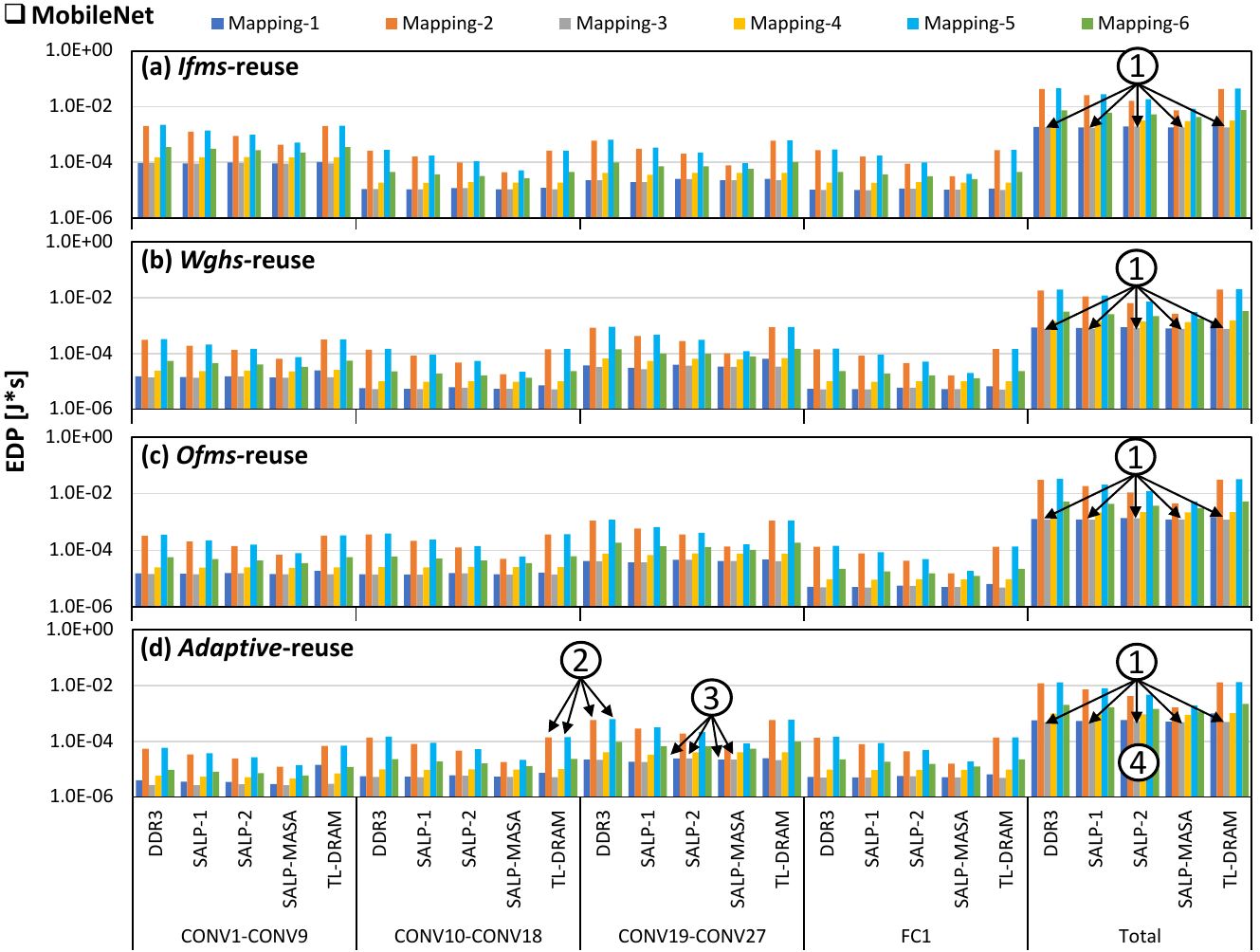}
\vspace{-0.2cm}
\caption{EDP of different DRAM mapping policies for MobileNet across different DRAM architectures (i.e., DDR3, SALP-1, SALP-2, SALP-MASA, and TL-DRAM), as well as different data partitioning and scheduling schemes: (a) \textit{ifms}-reuse, (b) \textit{wghs}-reuse, (c) \textit{ofms}-reuse, and (d) adaptive-reuse.}
\label{Fig_Result_MobileNet}
\vspace{-0.3cm}
\end{figure*}

\begin{figure*}[t]
\centering
\includegraphics[width=0.8\linewidth]{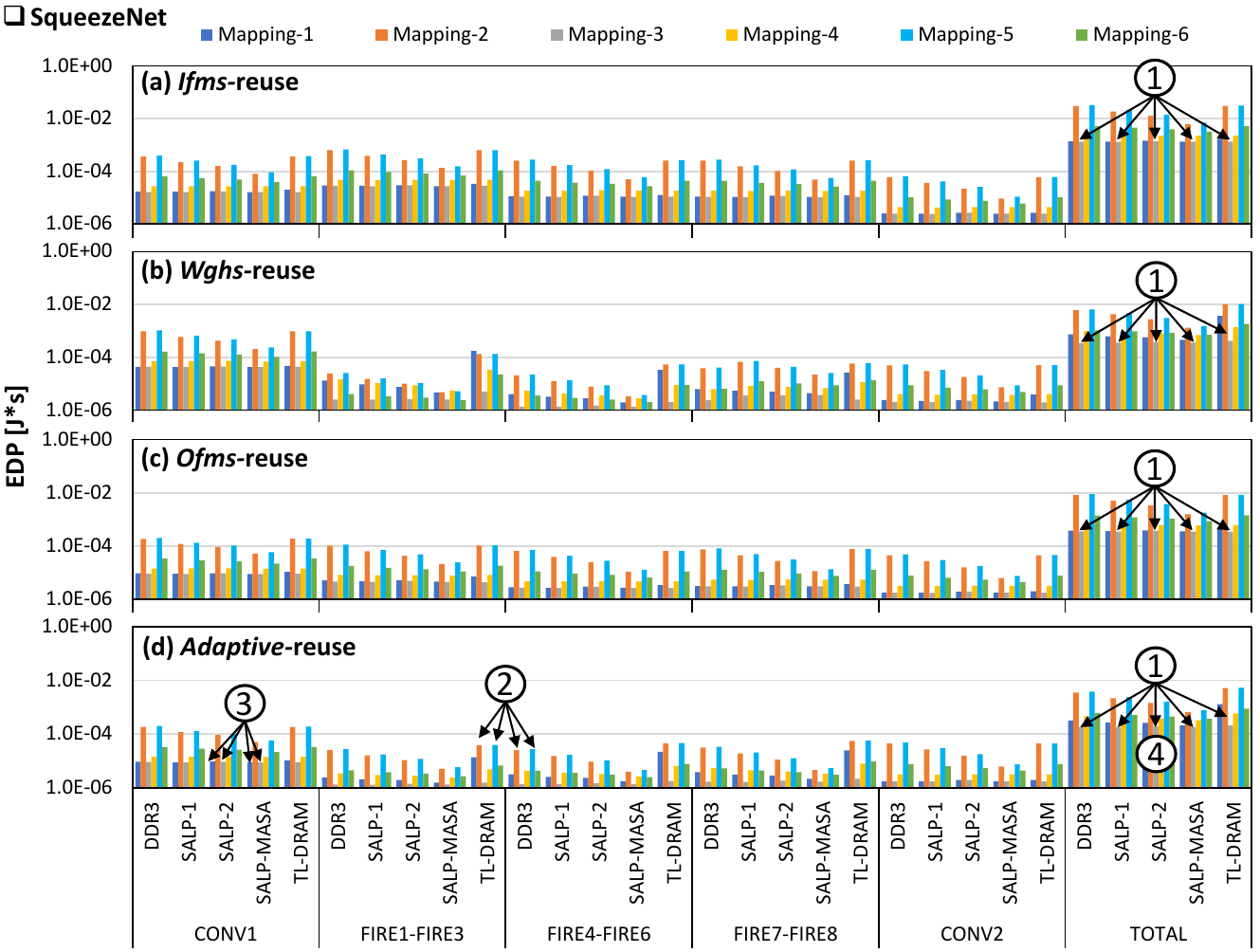}
\vspace{-0.3cm}
\caption{EDP of different DRAM mapping policies for SqueezeNet across different DRAM architectures (i.e., DDR3, SALP-1, SALP-2, SALP-MASA, and TL-DRAM), as well as different data partitioning and scheduling schemes: (a) \textit{ifms}-reuse, (b) \textit{wghs}-reuse, (c) \textit{ofms}-reuse, and (d) adaptive-reuse.}
\label{Fig_Result_SqueezeNet}
\vspace{-0.5cm}
\end{figure*}

\begin{figure*}[t]
\centering
\includegraphics[width=0.78\linewidth]{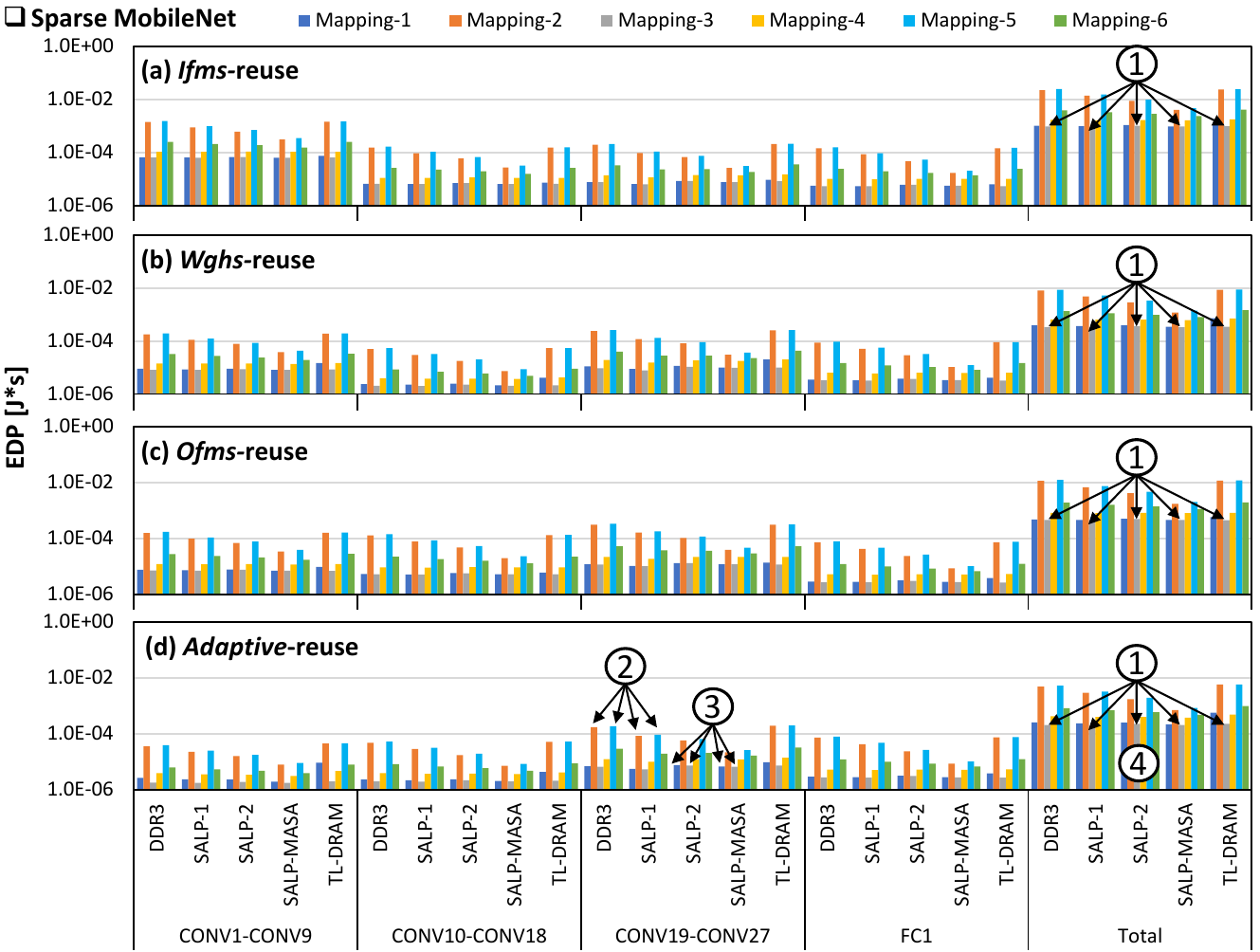}
\vspace{-0.3cm}
\caption{EDP of different DRAM mapping policies for the Sparse MobileNet across different DRAM architectures (i.e., DDR3, SALP-1, SALP-2, SALP-MASA, and TL-DRAM), as well as different data partitioning and scheduling schemes: (a) \textit{ifms}-reuse, (b) \textit{wghs}-reuse, (c) \textit{ofms}-reuse, and (d) adaptive-reuse.}
\label{Fig_Result_SparseMobileNet}
\vspace{-0.5cm}
\end{figure*}

We evaluate the impact of different DRAM mapping policies on the performance of a CNN accelerator across different DRAM architectures, data partitioning, and scheduling schemes. 
The experimental results are presented in Fig.~\ref{Fig_Result_AlexNet}-\ref{Fig_Result_SparseMobileNet} for the AlexNet, VGG-16, MobileNet, SqueezeNet, and Sparse MobileNet, respectively.

\vspace{-0.3cm}
\subsection{Comparisons of Different DRAM Data Mapping Policies}
\label{Chap_Results_CompareDRAMmap}

\textbf{Observation-\circled{1}:}
Our proposed DRAM data mapping policy (i.e., Mapping-3) achieves the lowest EDP across different layers of the network, different DRAM architectures, different scheduling schemes, and different networks.
It indicates that our proposed mapping is the most effective DRAM data mapping policy as it always achieves the smallest EDP for each layer of networks across different possible design settings, thereby meeting the optimization objective described in Section~\ref{Sec_PENDRAM_Model}. 
According to Table~\ref{Table_DRAMmaps}, our mapping policy (Mapping-3) \textit{orderly prioritizes} mapping the data to (1) different columns in the same row, which leads to row buffer hits in DDR3, SALPs, and TL-DRAM; (2) different banks in the same chip, which exploits bank-level parallelism in DDR3, SALPs, and TL-DRAM; (3) different subarrays in the same bank with priority mapping to the near segment rows, which exploits subarray-level parallelism in SALPs and near segment accesses in TL-DRAM, but leads to row buffer conflicts in DDR3; and (4) different rows in the same subarray with priority mapping to the near segment rows, which exploits near segment accesses in TL-DRAM but leads to row buffer conflicts in DDR3 and SALPs.
Following are detailed EDP improvements achieved by our DRAM data mapping policy (i.e., Mapping-3) as compared to other mapping policies.
\begin{itemize}[leftmargin=*]
    \item For the AlexNet, our mapping improves the EDP by up to 96\% in DDR3, 94\% in SALP-1, 88\% in SALP-2, 73\% in SALP-MASA, and 96\% in TL-DRAM. 
    \item For the VGG-16, our mapping improves the EDP by up to 96\% in DDR3, 94\% in SALP-1, 89\% in SALP-2, 77\% in SALP-MASA, and 96\% in TL-DRAM.
    \item For the MobileNet, our mapping improves the EDP by up to 96\% in DDR3, 94\% in SALP-1, 89\% in SALP-2, 79\% in SALP-MASA, and 95\% in TL-DRAM.
    \item In the SqueezeNet, our mapping improves the EDP by up to 95\% in DDR3, 93\% in SALP-1, 90\% in SALP-2, 81\% in SALP-MASA, and 95\% in TL-DRAM. 
    \item For the Sparse MobileNet, our mapping improves the EDP by up to 96\% in DDR3, 94\% in SALP-1, 89\% in SALP-2, 79\% in SALP-MASA, and 95\% in TL-DRAM. 
\end{itemize}

\textit{These results prove that our proposed mapping policy is the generalized DRAM data mapping policy that offers the lowest EDP for different design settings.}
Moreover, different DRAM access scheduling schemes can also benefit from our DRAM data mapping policy, so that the CNN accelerators that employ different scheduling schemes can optimize their DRAM access latency and energy.

\smallskip
\textbf{Observation-\circled{2}:} 
Mapping-2 and Mapping-5 obtain worse EDP values across different layers of the network, different DRAM architectures, and different scheduling schemes, than other mapping policies.
The reason is that, Mapping-2 and Mapping-5 prioritize mapping the data across different subarrays in the same bank, hence exploiting subarray-level parallelism in SALPs, but leading to row buffer conflicts in DDR3 and may lead to far segment accesses in TL-DRAM. 
Consequently, these mapping policies incur higher EDP values as compared to other mapping policies that mainly exploit row buffer hits (i.e., Mapping-1 and Mapping-3) and bank-level parallelism (i.e., Mapping-4 and Mapping-6). 

\smallskip
\textbf{Observation-\circled{3}:} 
Mapping-1 and Mapping-3 obtain comparable EDP values across different layers of the network, different DRAM architectures, and different scheduling schemes. 
The reason is that, Mapping-1 and Mapping-3 prioritize mapping the data across different columns in the same row, which leads to row buffer hits in DDR3, SALPs, and TL-DRAM.
The difference between these mapping policies comes when Mapping-1 prioritizes exploiting subarray-level parallelism over bank-level parallelism, while Mapping-3 is the opposite.
From observation, bank-level parallelism incurs lower latency and energy than subarray-level parallelism as shown in Fig.~\ref{Fig_HitMissConflict}, thereby leading to lower EDP values for Mapping-3 (i.e., our DRAM data mapping policy). 

\smallskip
\textbf{Observation-\circled{4}:} 
The adaptive-reuse scheduling scheme offers the lowest EDP for each layer of networks and the lowest total EDP for the given network models, as compared to other scheduling schemes. 
The reason is that, for each layer of networks, the adaptive-reuse scheduling scheme prioritizes keeping the data type that has the highest reuse factor on-chip for multiple operations, thereby maximizing the benefits of data reuse from different scheduling schemes (i.e., either \textit{ifms}-reuse, \textit{wghs}-reuse, or \textit{ofms}-reuse scheduling scheme).

\vspace{-0.2cm}
\subsection{Comparisons of Employing Different DRAM Architectures}
\label{Sec_Results_CompareDRAMarch}

In general, we observe that employing the SALP architectures (i.e., SALP-1, SALP-2, or SALP-MASA) can improve the EDP as compared to DDR3, across different networks. 
For instance, if we consider an adaptive-reuse scheduling scheme, the SALP architectures achieve EDP improvements by up to 88\% for AlexNet, by up to 87\% for VGG-16, by up to 86\% for MobileNet, by up to 81\% for SqueezeNet, and by up to 85\% for Sparse MobileNet. 
These EDP improvements are most notable in Mapping-2 and Mapping-5 as these mapping policies prioritize mapping data across subarrays in the same bank, thus exploiting subarray-level parallelism in SALP architectures but leading to row buffer conflicts in DDR3. 
Likewise, employing the TL-DRAM can also improve the EDP as compared to DDR3 across different networks.
For instance, if we consider an adaptive-reuse scheduling scheme, TL-DRAM achieves EDP improvements by up to 4\% for all investigated networks (i.e., AlexNet, VGG-16, MobileNet, SqueezeNet, and Sparse MobileNet). 
These EDP improvements are most notable in Mapping-5 as this mapping policy prioritizes mapping data across subarrays in the same bank, thereby exploiting near-segment row accesses in TL-DRAM but leading to row buffer conflicts in DDR3.

\textit{Although these mapping policies provide improvements in novel DRAM architectures (i.e., SALPs and TL-DRAM), their EDP values for DRAM accesses are still higher than the EDP values achieved by our DRAM data mapping policy (i.e., Mapping-3) across different scheduling schemes.}
Therefore, these results lead to several observation points as described in the following. 

\begin{itemize}[leftmargin=*]
    \item The EDP of employing different DRAM architectures may be different because of different DRAM access latency and energy profiles.
    \item Employing SALP architectures or TL-DRAM is beneficial for improving the energy efficiency of DRAM accesses in CNN accelerators, as long as an efficient mapping policy (i.e., our DRAM data mapping) is employed to achieve the lowest EDP of DRAM accesses.
    \item Since the internal organization of all DRAM architectures is similar (i.e., it is composed of channel, rank, chip, bank, subarray, row, and column), our PENDRAM methodology can be employed for all DRAM architectures to achieve energy-efficient processing of convolutional neural networks in CNN accelerators.
    \vspace{-0.2cm}
\end{itemize}

\vspace{-0.2cm}
\section{Conclusion}
\label{Sec_Conclusion}
\vspace{-0.1cm}

We present the PENDRAM methodology which employs a generalized DRAM data mapping policy that always offers the lowest EDP of DRAM accesses for enabling high-performance and energy-efficient CNN accelerators.
It is proven through an extensive DSE that evaluates the EDP values of different mapping policies across different DRAM architectures, as well as different data partitioning and scheduling schemes. 
Therefore, this work enables energy-efficient CNN accelerator designs and improves the DRAM access latency and energy for the existing CNN accelerators.

\vspace{-0.2cm}
\section*{Acknowledgment}
\label{Ack}
\vspace{-0.1cm}

This work was partially supported by the NYUAD Center for Artificial Intelligence and Robotics (CAIR), funded by Tamkeen under the NYUAD Research Institute Award CG010.

\vspace{-0.2cm}
\bibliographystyle{IEEEtran}
\bibliography{bibliography.bib}


\begin{IEEEbiography}[{\includegraphics[width=1in,height=1.25in,clip,keepaspectratio]{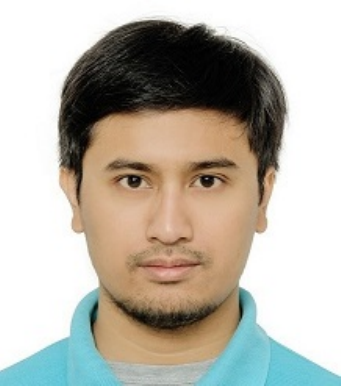}}]{Rachmad Vidya Wicaksana Putra}
(Member, IEEE) 
received B.Sc. on Electrical Engineering and M.Sc. on Electronics Engineering, both from Bandung Institute of Technology (ITB), Indonesia. 
He did PhD research at the Institute of Computer Engineering, Technische Universit\"at Wien (TU Wien), Vienna, Austria, during which he received multiple HiPEAC Paper Awards for his first-authored papers and an ACM Showcase. 
In academia, he has worked as a teaching assistant at the Electrical Engineering ITB, a research assistant at Microelectronics Center ITB, and a project research assistant at the Institute of Computer Engineering, TU Wien. 
In industry, he also experienced working as an FPGA engineer at the PT. Fusi Global Teknologi, Indonesia, and at the TriLite Technologies GmbH, Austria. 
Currently, he is a Research Team Leader at the eBrain Lab, New York University (NYU) Abu Dhabi, Abu Dhabi, UAE.
His research interests include brain-inspired \& neuromorphic computing, computer architecture, integrated circuits \& VLSI, emerging technologies, system-on-chip design, robust \& energy-efficient computing, embedded AI, and electronic design automation.
\end{IEEEbiography}

\begin{IEEEbiography}[{\includegraphics[width=1in,height=1.25in,clip,keepaspectratio]{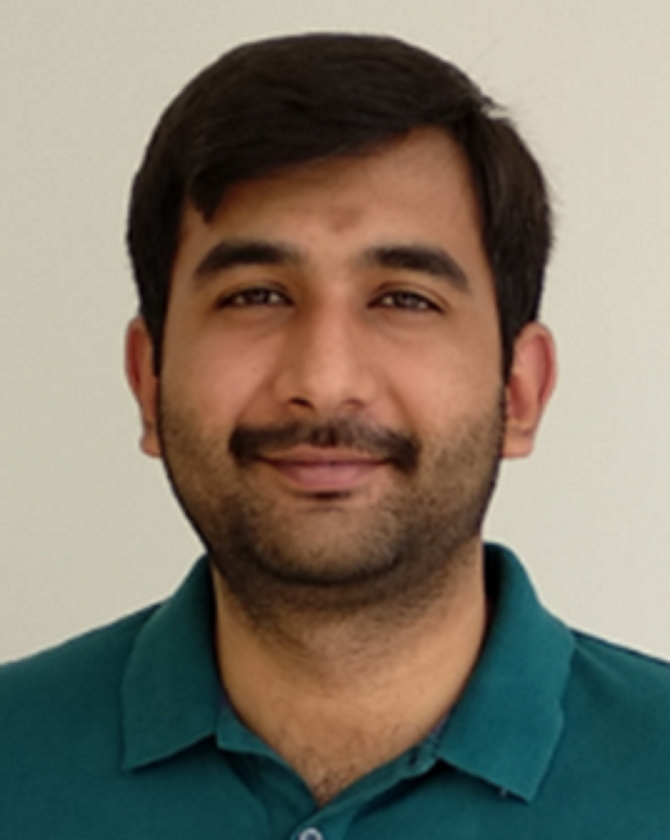}}]{Muhammad Abdullah Hanif}
(Member, IEEE) received the B.Sc. degree in electronic engineering from the Ghulam Ishaq Khan Institute of Engineering Sciences and Technology (GIKI), Pakistan, the M.Sc. degree in electrical engineering with a specialization in digital systems and signal processing from the School of Electrical Engineering and Computer Science, National University of Sciences and Technology (NUST), Islamabad, Pakistan, and the Ph.D. degree in computer engineering from the Vienna University of Technology (TU Wien), Austria. 
He is currently a Research Group Leader with New York University (NYU) Abu Dhabi, United Arab Emirates. His research interests include brain-inspired computing, machine learning, approximate computing, computer architecture, energy-efficient design, robust computing, system-on-chip design, and emerging technologies.
\end{IEEEbiography}

\begin{IEEEbiography}[{\includegraphics[width=1in,height=1.25in,clip,keepaspectratio]{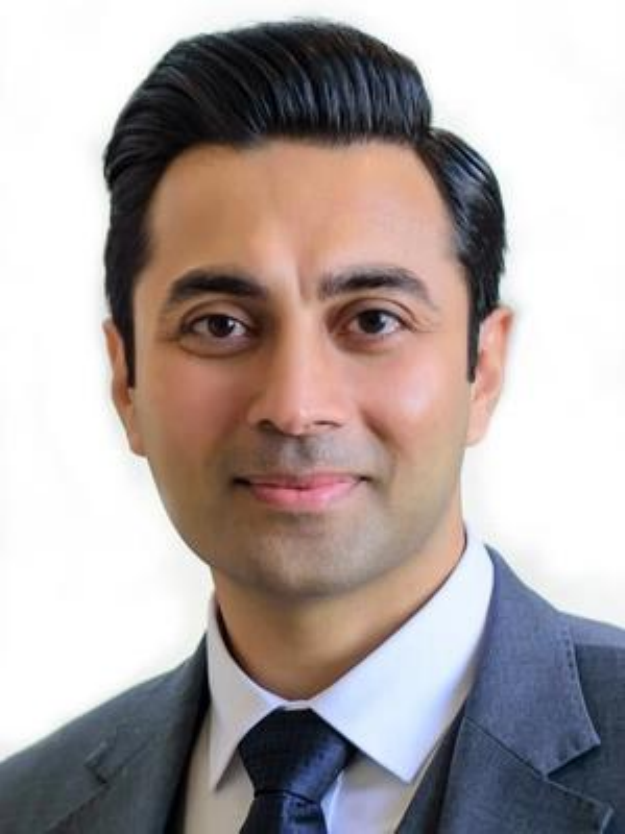}}]{Muhammad Shafique}
(Senior Member, IEEE)
received the Ph.D. degree in computer science from
Karlsruhe Institute of Technology (KIT), Karlsruhe,
Germany, in 2011.

He established and led a highly recognized research group at KIT for several years as well as conducted impactful collaborative R\&D activities across the globe. 
In 2016, he joined as a Full Professor in computer architecture and robust, energy-efficient technologies at the Faculty of Informatics,
Institute of Computer Engineering, Technische Universit\"at Wien (TU Wien), Vienna, Austria. 
Since 2020, he has been with New York University (NYU) Abu Dhabi, Abu Dhabi, UAE, where he is currently a Full Professor and the Director of eBrain Lab. 
He is a Global Network Professor with Tandon School of Engineering, NYU, New York, NY, USA. 
He is also a Co-PI/Investigator with multiple NYUAD Centers, including Center of Artificial Intelligence and Robotics (CAIR), Center of Cyber Security (CCS), Center for InTeractIng urban nEtworkS (CITIES), and Center for Quantum and Topological Systems (CQTS). 
His research interests include AI \& machine learning hardware and system-level design, brain-inspired computing, quantum machine learning, cognitive autonomous systems, wearable healthcare, energy-efficient systems, robust computing, hardware security, emerging technologies, FPGAs, MPSoCs, and embedded systems. 
His research has a special focus on cross-layer analysis, modeling, design, and optimization of computing and memory systems. 
The researched technologies and tools are deployed in application use cases from Internet-of-Things (IoT), smart cyber–physical systems (CPSs), and ICT for development (ICT4D) domains. 
He has given several keynotes, invited talks, and tutorials, as well as organized many special sessions at premier venues. 
He has served as the PC Chair, the General Chair, the Track Chair, and a PC member for several prestigious IEEE/ACM conferences. 
He holds one U.S. patent and has (co-)authored 6 books, 10+ book chapters, 350+ papers in premier journals and conferences, and 100+ archive articles.

Dr. Shafique received the 2015 ACM/SIGDA Outstanding New Faculty
Award, the AI 2000 Chip Technology Most Influential Scholar Award in
2020, 2022, and 2023, the ASPIRE AARE Research Excellence Award in 2021,
six gold medals, and several best paper awards and nominations at prestigious conferences. 
He is a senior member of the IEEE Signal Processing Society (SPS) and a member of the ACM, SIGARCH, SIGDA, SIGBED, and HIPEAC.
\end{IEEEbiography}

\end{document}